# Is room temperature ferromagnetism possible in K-doped $SnO_2$?

**S. K. Srivastava*[1], P. Lejay[1], B. Barbara[1] S. Pailhès[2, 3], V. Madigou[4] and G. Bouzerar[1, 5]**

[1]Institut Néel, CNRS Grenoble and Université Joseph Fourier, 38042 Grenoble, Cedex 9, France

[2]University Lyon 1 & CNRS, 69622 Villeurbanne, France

[3]Laboratoire Léon Brillouin (LLB), CEA/CNRS Saclay, 91191 Gif-sur-Yvette, France

[4]University Sud Toulon-Var, IM2NP-CNRS (UMR 6242), Bâtiment R, BP 132, F-83957 La Garde Cedex, France

[5]Jacobs University Bremen School of Engineering and Science, Campus Ring 1, D-28759 Bremen, Germany

## Abstract

Ab initio studies have theoretically predicted room temperature ferromagnetism in crystalline $SnO_2$, $ZrO_2$ and $TiO_2$ doped with non magnetic element from the 1A column as K and Na. Our purpose is to address experimentally the possibility of magnetism in both $Sn_{1-x}K_xO_2$ and $Sn_{1-x}Ca_xO_2$ compounds. The samples have been prepared using *equilibrium* methods of standard solid state route. Our study has shown that both $Sn_{1-x}Ca_xO_2$ and $Sn_{1-x}K_xO_2$ structure is thermodynamically unstable and leads to a phase separation, as shown by X-ray diffraction and detailed micro-structural analyses with high resolution transmission electron microscopy (TEM). In particular, the crystalline $SnO_2$ grains are surrounded by K-based amorphous phase. In contrast to Ca: $SnO_2$ samples we have obtained a magnetic phase in K: SnO2 ones, but no long range ferromagnetic order. The K: $SnO_2$ samples exhibit a moments of the order of 0.2 $\mu_B$/K /ion, in contrast to ab-initio calculations which predict 3$\mu_B$, where K atoms are on the Sn crystallographic site. The apparent contradictions between our experiments and first principle studies are discussed.

* **E-mail:** sandeep.srivastava@grenoble.cnrs.fr

In recent years, spintronics has become one of the most important fields of research as demonstrated by the huge existing literature. A frenetic race which involves both fundamental research scientists and industrial partners has been recently triggered to discover and elaborate a new family of materials for spintronics: the ferromagnetic semi-conductors. The interest in such materials consists in the use of new devices where spin degrees of freedom carry information in order to reduce electrical consumption, allow non-volatility, store and manipulate data, beyond room temperature. The search for candidates with room temperature ferromagnetism is really intense but, in most cases, materials are not very well controlled. In particular, it has been argued that ferromagnetism can be observed in materials, referred to as $d^0$ or intrinsic ferromagnets, which do not contain magnetic impurities. Coey *et a*l. [1] reported $d^0$ ferromagnetism above room temperature in un-doped $HfO_2$ thin films fabricated by conventional out of equilibrium pulsed-laser deposition. Such unexpected ferromagnetism has also been reported in CaO, ZnO, $CaB_6$ and $ZrO_2$ [2-5]. But contradictory experiments are also published concluding to the absence of magnetic moments in same systems (see e.g. Ref [6] for $HfO_2$) showing that experimental proofs of intrinsic $d^0$ ferromagnetism are controversial. It appears clearly that non-controlled preparation methods or/and history of samples make difficult to stabilize equilibrium systems allowing to study intrinsic $d^0$ magnetism. From a theoretical point of view, it has been shown that point defects as cation vacancies may be at the origin of the magnetism of these materials [2, 7-9]. A vacancy induces local magnetic moments on the neighboring oxygen atoms which then interact with extended exchange couplings. A progress has been done recently in the understanding of the $d^0$ ferromagnetism in the framework of a minimal model which includes disorder and electron-electron correlation effects on equal footing [10]. It was predicted that high Curie temperatures ($T_C$) could be reached for realistic physical parameters. To circumvent the difficulties of defect control, an alternative way, consisting in the substitution of the cation $A^{4+}$ in dioxides such as $AO_2$ ($A$=Ti, Zr, or Hf) by a monovalent cation of the group 1A, was proposed. A recent ab initio study predicted, for $K^+$ or $Na^+$ doped $ZrO_2$, ferromagnetism with $T_C$ far beyond room temperature [5].

$SnO_2$ is a wide band-gap material, used as a transparent conduction electrode in flat panel display and solar cells [11]. It has a rutile structure with distorted octahedral coordination. The possibility of ferromagnetism in $SnO_2$ by doping with nonmagnetic K & Ca element has not been investigated experimentally yet. $Sn_{1-x}K_xO_2$ and $Sn_{1-x}Ca_xO_2$ compounds ($0 \le x \le 0.08$) were prepared by standard solid state route method by using high-purity $SnO_2$ (purity, 99.996%), $K_2CO_3$ (99.99 %) and $CaCO_3$ (purity, 99.995%) compounds. The maximum amount of any kind of trace magnetic impurities in the starting materials was found to be less than 0.9 % ppm as mentioned by the supplier ICP chemical analyses report. The final annealing in pellet form was carried out at $500^0$C for 20 hrs in air. Slow scan powder X-Ray diffraction patterns were collected by using Philips XRD machine with $CuK_\alpha$ radiation. Micro-structural images and chemical analyses have been carried out by a transmission electron microscope (TEM: Tecnai $G^2$ operating at 200 kV) equipped with an energy dispersive spectrometer (EDS). Magnetization measurements as a function of magnetic field H and temperature T were carried out using a commercial SQUID magnetometer (Quantum Design, MPMS).

The X-Ray diffraction patterns for K: SnO2 samples along with one typical XRD pattern for Ca: SnO2 are shown in Figure 1. All the diffractions peaks could be indexed on the basis of the tetragonal rutile type-structure. No extra diffraction peaks detected showing that no crystalline parasitic phases are present in the samples within the limit of XRD. However, we did not observe any significant change in the lattice parameters obtained from the refinement of XRD patterns with Fullprof program. The lattice parameters for pure $SnO_2$ was found to be a = b = 4.7385 Å and c = 3.1871 Å, and are comparable to those reported by Duan *et al.* [12]. This observation strongly indicates that no solid solution is formed i.e. that $K^+$ ions do not enter grain cores. This can be understood by the large difference between the ionic radii of the 6-coordinate $Sn^{4+}$ (0.69Å) and that of $K^+$ (1.38 Å) & $Ca^+$ (1 Å). In the case of $K^+$, significant line broadening appears at largest concentrations.

To understand the origin of such broadening at the microstructure level we have performed observations by TEM coupled with EDS. One typical TEM image of K: SnO2 with 8 atm. % of K is shown in Figure 2(a). The electron diffraction patterns (line positions and widths) obtained on the grains can be indexed by the rutile type structure of $SnO_2$. The Figure 2(b) shows the selected area diffraction pattern of grains with a [101] zone axis. The high-resolution transmission electron microscope (HRTEM) image recorded at different locations shows a (110) in-plane preferential orientation of nanometer size grains as can be seen in Figure 2(c). The EDS measurement using the TEM facility was carried out and it showed, on each grain, the presence of K concentration close to the sample nominal concentration of 8% (average value). Thus from the X-ray and TEM analyses, we can conclude that the compounds consist of crystalline $SnO_2$ grain-cores coated with a K-based amorphous phase.

The magnetic measurements for all samples were done with utmost care and repeated three times with different pieces of samples to guarantee the reproducibility of results. The magnetic properties of all the starting compounds; $SnO_2$, $CaCO_3$ and $K_2CO_3$ have been checked and they clearly exhibit diamagnetic behavior. The M-H measurements of all Ca: SnO2 (up to 8%) compounds exhibit diamagnetic behavior down to 2.5 K (as shown in the inset of Figure 3a). Thus, there is no magnetism induced by Ca doping. However, K: SnO2 samples were found to be very interesting. The M-H measurements show that up to 4% (small X-ray line-widths) K-doped compounds exhibit diamagnetic behavior down to 2.5K, as shown in Figure 3 (a). However, the 6% and 8% compounds (larger X-ray line-widths) are found to be magnetic and their magnetization was found to increase proportionally to the K concentration as shown in Figure 3 (b) and 3(c). It approaches saturation with the values ~ 0.010 and ~ $0.019\mu_B$/K-ion at 2.5 K, respectively. The magnetic susceptibility decreases with temperature in the whole range of measurements and follows a Curie-Weiss law with paramagnetic Curie temperatures and effective moments $\theta_p$ =7 K, $\mu_{eff} = 0.20\mu_B$ for the 6%, and $\theta_p$ =11 K, $\mu_{eff} \sim 0.25\mu_B$ for the 8% of K sample. A low temperature upturn of the reciprocal susceptibility increasing with K-doping (Figure 3d) suggests that antiferromagnetic (AFM) interactions compete with dominant ferromagnetic (FM) interactions responsible for positive $\theta_p$. Such competing interactions may lead to spin-canted ferromagnetic blocks or to antiferromagnetically coupled ferromagnetic blocks, explaining why the M-H curves give low-temperature magnetic saturation of the order of 0.01-0.02 $\mu_B$ which is ~10 times smaller than the effective paramagnetic moment $\mu_{eff} \sim 0.22 \pm 0.02$ $\mu_B$. Here, we must emphasis that we have always observed diamagnetism for $SnO_2$, Ca doped and low K doped $SnO_2$ samples. The introduction of K $\geq$ 0.06% only induces a paramagnetic moments of ~ 0.22 ± 0.02 $\mu_B$/K together with a grain size reduction (larger X-ray line-widths). According to TEM and XRD data previously discussed, we attribute theses effects to the existence of core-shell nano-particles, in which surface magnetism is induced by $K^+$ ions. Thus we would expect an increase of the magnetic moment when the ratio volume/surface decreases. This is indeed observed according to X-ray measurements as the line-widths & the average grain size is larger (200 nm) for x= 0.06 than for x= 0.08 (100 nm)).

Let us now discuss the existing first principle studies based on pseudo-potential LDA method for K and Ca doped $SnO_2$ [13]. In Ref. 13 it has been predicted a large magnetic moment of $3\mu_B$, comparable to that obtained in other samples of K, and Na doped $ZrO_2$ [5]. In the ab initio studies one assumes a real substitution of Sn by Ca or K. The essential point is that in ab initio studies thermodynamics is ignored. Experimentally we have seen that K/Ca in $SnO_2$ leads to a phase separation and the substitution is not thermodynamically possible. However, the formation of amorphous-K out of equilibrium ion-shells induces some magnetism. Thus it would be interesting to analyze the occurrence of phase separation within first principle study. Note however that no magnetism was found in Ca doped sample which agrees qualitatively with our experimental results.

To conclude, we have prepared $SnO_2$ :( K, Ca) compounds by solid states route method. The X-ray diffraction and the detailed micro structural analyses with high resolution transmission electron microscopy (TEM) provide the evidence of a core-shell structure with crystalline $SnO_2$ grains surrounded by K-based amorphous phase. In contrast to Ca: SnO2 samples we have obtained a magnetic phase in K: $SnO_2$ ones, but no long range ferromagnetic order. The K: $SnO_2$ compounds exhibit a moments of the order of 0.2 $\mu_B$/K /ion, in contrast to ab-initio calculations which predict 3$\mu_B$, where K atoms substitute the Sn ones. The origin of observed magnetism in K: $SnO_2$ compounds are attributed to the existence of core-shell nano-particles, in which surface magnetism is induced by $K^+$ based amorphous phase. The difference between experiments and theory comes from the fact that the Rutile crystal structure used in ab initio calculations is unstable upon K-doping, and therefore does not correspond to the real one. One of the messages of this paper is that first principle calculations should be preceded by thermodynamically calculations showing what is the most stable crystal structure in the presence of non-magnetic doping ions (K in the present case).

## References:

[1]M. Venkatesan, C.B. Fitzgerald, J.M.D. Coey, Nature **430,** 630 (2004).

[2]J.Osorio-Guillen, S. Lany, S.V. Barabash, A. Zunger, Phys. Rev. Lett. **96,** 107203 (2006).

[3]H. Pan, J.B. Yi, L. Shen, R.Q. Wu, J.H. Yang, J.Y. Lin, Y.P. Feng, J. Ding, L.H. Van, J.H. Yin,, Phys. Rev. Lett. **99,** 127201 (2007).

[4]D. P. Young, D. Hall, M. E. Torelli, Z. Fisk, J. L. Sarrao, J. D. Thompson, H.-R. Ott, S. B. Oseroff, R. G. Goodrich, R. Zysler, Nature **397,** 412 (1999).

[5]F. Maca, J. Kudrnovsky, V. Drchal, G. Bouzerar, Appl. Phys. Lett. **92,** 212503 (2008).

[6]N Hadacek, A Nosov, L Ranno, P Strobel and R-M Galera, J. Phys.: Condens. Matter **19**, 486206 (2007).

[7]C. Das Pemmaraju, S. Sanvito, Phys. Rev. Lett. **94,** 217205 (2005).

[8]A. M. Stoneham, A. P. Pathak, and R. H. Bartram, J. Phys. C **9,** 73 (1976).

[9]S. Elfimov, S. Yunoki, and G. A. Sawatzky, Phys. Rev. Lett. **89,** 216403 (2002).

[10]G. Bouzerar, T. Ziman, Phys. Rev. Lett. 96 207602 (2006).

[11]H. L. Hartnagel, A. L. Dawar, A. K. Jain and C. Jagadish, Semiconducting Transparent Thin Films, Bristol, 1995.

[12]L.B.Duan, G.H.Rao, J.Yu,Y.C.Wang, G.Y.Liu, J.K.Liang, J. Appl. Phys **101,** 063917 (2007).

[13]W. Zhou, L. Liu, P. Wu, J. of Magnetism and Magnetic Materials **321,** 3356 (2009).

## Figure Captions:

**Figure 1:** XRD patterns of K:$SnO_2$ samples and one representative XRD pattern for Ca:$SnO_2$ sample.

**Figure 2: (**a) Normal resolution TEM image of K:$SnO_2$ with 8 at. % of K (b) Electron diffraction of selected grain oriented along the [101] direction (c) High Resolution TEM image showing the (110) and (101) planes.

**Figure 3:** M-H loops recorded for K:$SnO_2$ with x=0, 0.02, 0.04, 0.06, 0.08 % of K samples. Inset shows the M-H loops recorded for Ca doped samples (b) and (c) M-H loops recorded for x=0.06 and 0.08 % K doped samples at various temperature respectively (d) Curie-Weiss law fitting with $\theta_p$ extrapolations (arrows).

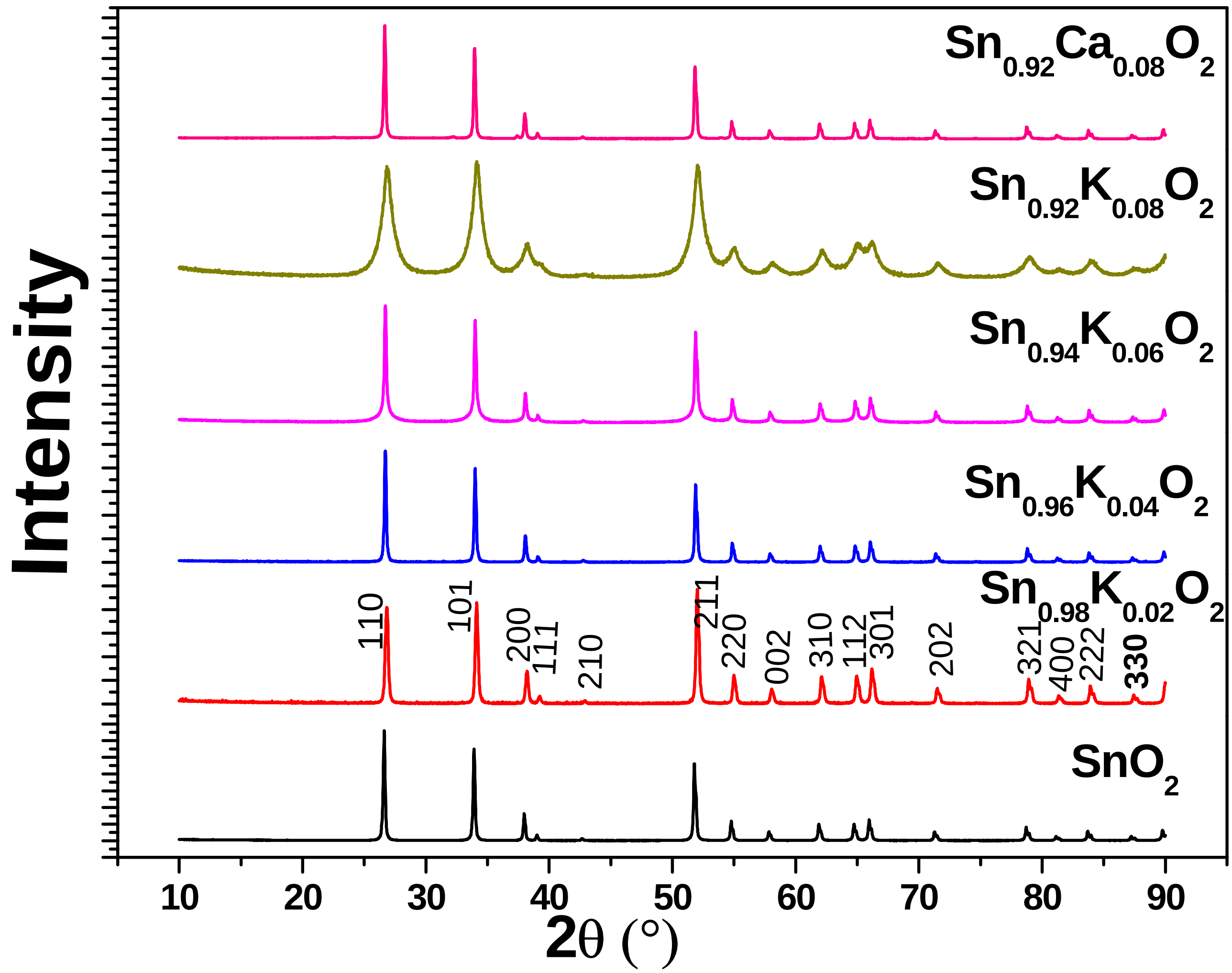

**Figure 1:** XRD patterns of K:$SnO_2$ samples and one representative XRD pattern for Ca:$SnO_2$ sample.

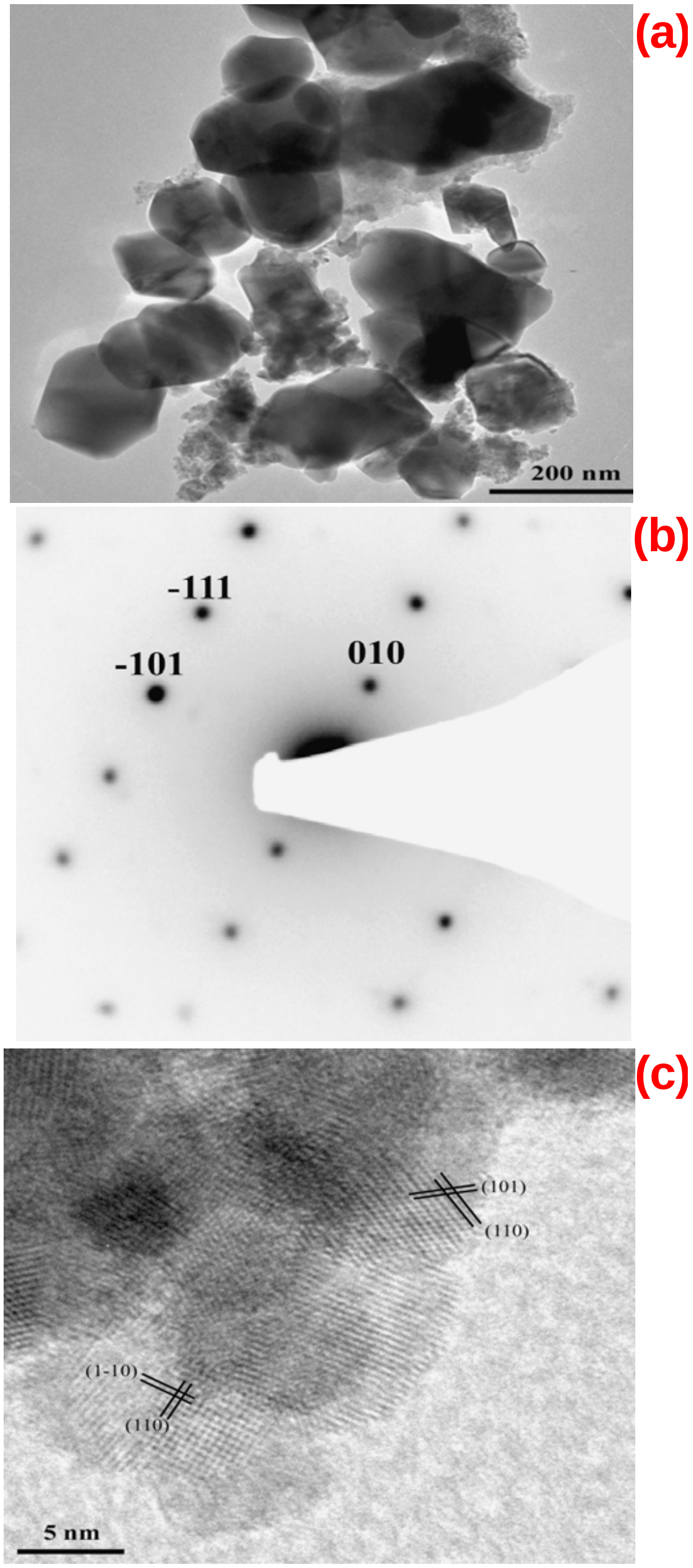

**Figure 2:** (a) Normal resolution TEM image of K:$SnO_2$ with 8 at. % of K (b) Electron diffraction of selected grain oriented along the [101] direction (c) High Resolution TEM image showing the (110) and (101) planes.

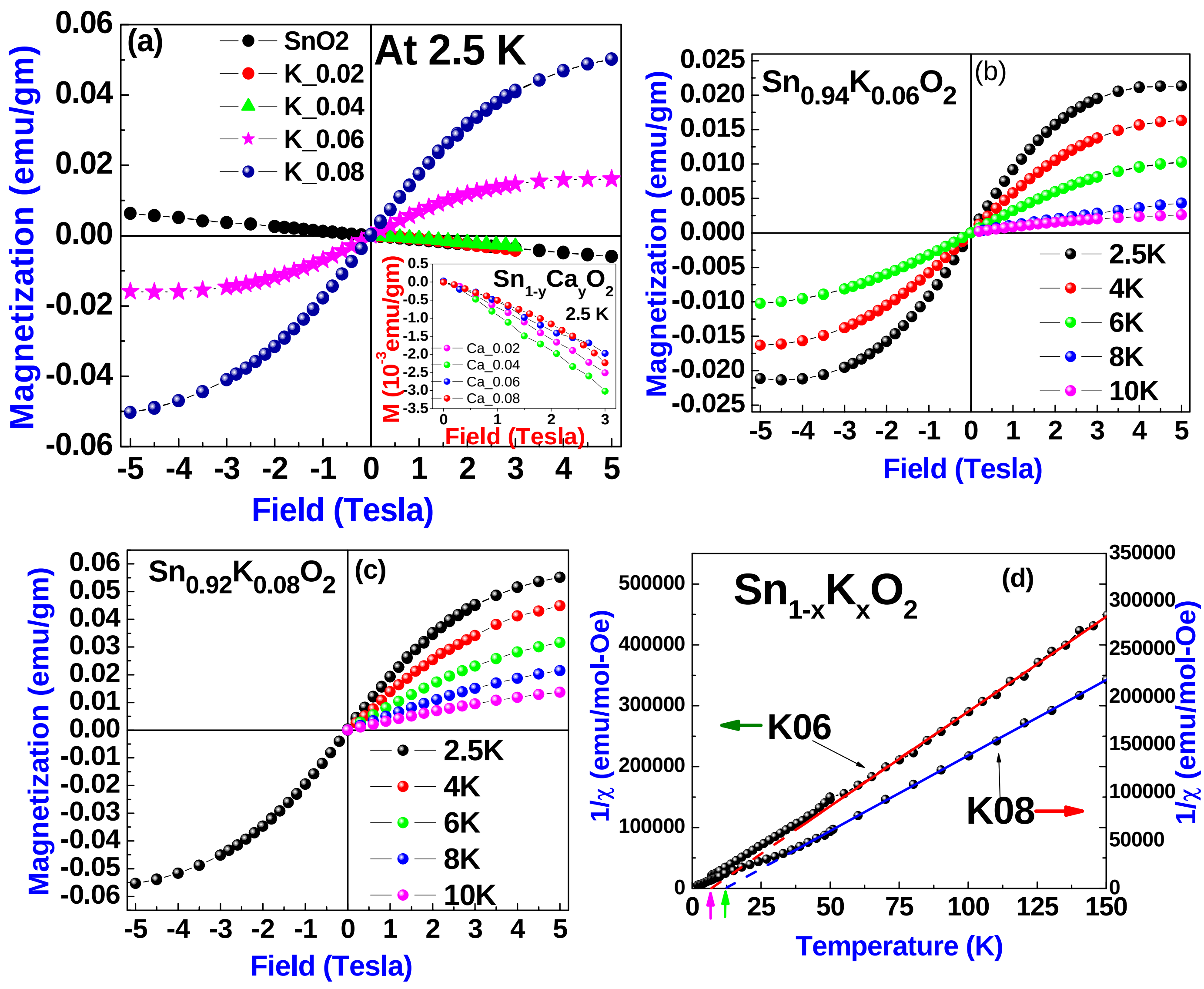

**Figure 3:** M-H loops recorded for K:$SnO_2$ with x=0, 0.02, 0.04, 0.06, 0.08 % of K samples. Inset shows the M-H loops recorded for Ca doped samples (b) and (c) M-H loops recorded for x=0.06 and 0.08 % K doped samples at various temperature respectively (d) Curie-Weiss law fitting with $\theta_p$ extrapolations (arrows).